\documentclass[11pt,a4paper]{article}

\usepackage{amsmath,amssymb}     
\usepackage{color}
\usepackage{graphicx}
\usepackage{subfig}
\usepackage{cite}                
\usepackage{hyperref}            
\usepackage{multirow,makecell}
\usepackage{braket,bm}
\usepackage{cancel}
\RequirePackage{doi}
\usepackage{hyperref}
\usepackage{leftidx}
\usepackage[text={17cm,24.5cm},centering]{geometry}
\usepackage{textcomp}
\usepackage{wasysym}
\usepackage{mathtools}

\numberwithin{equation}{section}   

\def \be {\begin{equation}}
\def \ee {\end{equation}}
\def \ba {\begin{array}}
\def \ea {\end{array}}
\def \bea{\begin{eqnarray}}
\def \eea{\end{eqnarray}}

\def \G {\Gamma}
\def \d {\delta}
\def \D {\Delta}
\def \dg {\dagger}

\def \m {\mu}

\def \l {\lambda}

\def \s {\sigma}

\def \o {\omega}
\def \O {\Omega}

\def \mE {\mathcal E}

\def \mN {\mathcal N}

\def \mP {\mathcal P}
\def \mQ {\mathcal Q}
\def \mR {\mathcal R}

\def \mT {\mathcal T}

\def \mV {\mathcal V}

\def \lt {\left}
\def \rt {\right}

\def \inf {\infty}

\def \Re {{\textrm{Re}}}

\def \Tr {{\textrm{Tr}}}

\def \and {{\textrm{and}}}

\begin{document}
\begin{titlepage}
	\title{\textbf {Dynamics of charge imbalance resolved negativity after a local joining quench}}
    \author{Hui-Huang Chen$^{a}$\footnote{chenhh@jxnu.edu.cn}~, Zun-Xian Huang$^{a}$\footnote{zxhuang@jxnu.edu.cn}}
	\maketitle
	\underline{}
	\vspace{-12mm}
	\begin{center}
		{\it
             $^a$College of Physics and Communication Electronics, Jiangxi Normal University,\\ Nanchang 330022, China\\
		}
		\vspace{10mm}
	\end{center}
	\begin{abstract}
 In this paper, we consider the dynamics of charge imbalance resolved negativity after a local joining quench in the $1+1$ dimensional free complex boson CFT. In the first part, we study the local joining quench by applying conformal maps, obtaining analytical universal results. We first calculate the quench dynamics of charged logarithmic negativity. Then using the Fourier transformation, we obtain the charge imbalance resolved negativity. The total negativity can be recovered from the charge-resolved ones. In the second part, we test our CFT predictions against the underlying lattice model numerically. Finally, we explain our results based on the quasi-particle picture.
	\end{abstract}
\end{titlepage}
\thispagestyle{empty}
\newpage
\tableofcontents
\newpage
\section{Introduction}
\label{Intro}
\par Over the last two decades, various entanglement measures have been found and played crucial roles in relating diverse fields ranging from field theories and many-body physics to quantum gravity and black hole physics
\cite{Calabrese:2004eu,Calabrese:2009qy,Amico:2007ag,Calabrese:2014yza,Chen:2015usa,Guo:2015uwa,Coser:2015eba,Nishioka:2009un,Solodukhin:2011gn,Fang:2016ehk,He:2017lrg}. In field theories and many-body physics, entanglement is used for
characterizing different phase of matter. In quantum gravity and black hole physics, since Ryu and Takayanagi first proposed the notion of holographic entanglement entropy\cite{Ryu:2006bv}, a large number of studies have emerged, including holographic entanglement entropy \cite{Cai:2012sk,Cai:2012nm,FarajiAstaneh:2013oey,Nozaki:2013vta,Kim:2014qpa,Kim:2015rvu,Tsujimura:2021xmj,He:2014lfa} and holographic entanglement negativity
\cite{Chaturvedi:2016rcn,Jain:2017uhe,Jain:2017aqk,Jain:2018bai,Malvimat:2018txq,Afrasiar:2021hld}.
\par Among the aforesaid research progresses, entanglement entropy is the most successful entanglement measure to characterize the bipartite entanglement of a subsystem $A$ in a pure state. Given a system in a pure state, the reduced density matrix (RDM)  of a subsystem $A$ is defined by tracing out its complement $B$ as $\rho_A=\Tr_B\ket{\Psi}\bra{\Psi}$. The R\'enyi entropies are defined as
\be
S_n=\frac{1}{1-n}\log\Tr\rho_A^n.
\ee
Then we can obtain the Von Neumann entropy through the replica trick \cite{Calabrese:2004eu}
\be
S_1 \equiv\lim_{n\rightarrow 1}S_n=-\lim_{n\rightarrow 1}\frac{\partial}{\partial n}\Tr \rho_A^n=-\Tr(\rho_A\log\rho_A).
\ee
\par
When two subsystems $A_1$ and $A_2$ embedded in a larger system are not necessarily complementary to each other, $\rho_{A_1\cup A_2}$ is in general a mixed state. Entanglement entropy is no longer a good measure of entanglement, since it mixes classical and quantum correlations. A computable measure of the bipartite entanglement for a general mixed state is the entanglement negativity, which can be defined as \cite{Vidal:2002zz,Plenio:2005cwa}
\be
\mathcal{N} = \frac{\Tr |\rho_A^{T_2}|-1}{2},
\ee
where $\Tr |O|=\Tr \sqrt{O^{\dg}O}$ denotes the trace norm of the operator $O$ and $\rho_A^{T_2}$ is the partial transpose of RDM $\rho_A$ with respect to the degree of freedom of subsystem $A_2$. In detail, given $|e_i^1 \rangle$ and $| e_j^2\rangle$ being the basis of Hilbert spaces $\mathcal{H}_1$,$\mathcal{H}_2$ associated to $A_1$ and $A_2$, respectively, the partial transposition with respect to the degrees of freedom of subsystem $A_2$ is defined as
\be
\bra{e_i^{(1)}e_j^{(2)}}\rho_A^{T_2}\ket{e_k^{(1)}e_l^{(2)}}=\bra{e_i^{(1)}e_l^{(2)}}\rho_A\ket{e_k^{(1)}e_j^{(2)}}.
\ee
The R\'enyi negativity and R\'enyi logarithmic negativity  are defined as
\be
\mathcal{N}_n=\frac{N_n-1}{2},\quad
\mE_n =\log N_n,\quad \text{with}\quad
N_n=\Tr [(\rho_A^{T_2})^n],
\ee
from which the negativity and logarithmic negativity are obtained by using the analytic continuation of the even integer at $n_e =1$
\be\label{logNeg}
\mN=\lim_{n_e \to 1}  \mN_{n_e},\quad
\mE=\lim_{n_e \to 1}  \mE_{n_e}.
\ee
\par
\begin{figure}
\centering
\includegraphics[scale=0.8]{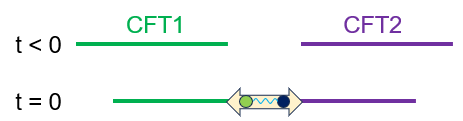}
\caption{Setup for local joining quench. Two separate CFTs are joined together at their endpoints at time $t=0$. Entangled pairs which carry entanglement information are quasi-particles generated at the joining point propagating through the system.}
\label{ljq}
\end{figure}

 Recently, the non-equilibrium evolution of isolated quantum system is one of the most active research area. The quantum quench is the simplest non-equilibrium setting \cite{Calabrese:2007rg}. Entanglement negativity after a quantum quench has been studied in field theories \cite{Calabrese:2006rx} and holographic theories\cite{Nozaki:2013wia,Jahn:2017xsg}. In this paper, we focus on the local joining quench which has been first studied in references \cite{Eisler:2007,Calabrese:2007mtj}. In particular, we will consider a local joining quench in a (1+1)-dimensional critical system \cite{Calabrese:2006rx,Calabrese:2007mtj}. As shown in Fig.~\ref{ljq}, the initial state can be represented by a line physically cut into two parts prepared in their own ground states at $t<0$ . We join the two parts together at their endpoints at $t=0$, then they evolve together.
\par A related quench setting is the so-called local operator quench, where initially a local operator was inserted at the origin. The increased value of entanglement entropy is related to the quantum dimension the corresponding operator \cite{Nozaki:2014hna}. For quasi-particle interpretation of local quench, the physic picture is observed in reference \cite{Nozaki:2014hna} in free field theory. For generic CFT, it has been proved in a following paper \cite{He:2014mwa}. Further, this picture has been proved in finite dimensional tensor category in \cite{Guo:2018lqq}.
\par An interesting question concerns the interplay between symmetries and entanglement\cite{Caputa:2013eka,Matsuura:2016qqu}. When a system has a global symmetry, the entanglement entropy and negativity will split into different symmetry sectors characterized by eigenvalues of some charge operator. The concept of symmetry-resolved entanglement particularly attracted much attention in both equilibrium and out-of-equilibrium cases\cite{Goldstein:2017bua,Cornfeld:2018wbg,
Bonsignori:2019naz,Murciano:2019wdl,Capizzi:2020jed,Murciano:2020lqq,Murciano:2020vgh,Horvath:2020vzs,Xavier:2018kqb,Parez:2021pgq,Murciano:2021djk,Calabrese:2021wvi,Ares:2022hdh}. In the context of mixed states, the dynamics of charge imbalance resolved negativity after a global quench in the free fermion and free boson models were investigated in \cite{Parez:2022xur} and \cite{Chen:2022gyy}, respectively. The dynamics of charge resolved entanglement and negativity after a local quench were studied in \cite{Feldman:2019upn} only for adjacent intervals. However, a thorough study of the charge imbalance resolved negativity after a local quench is still lacking.
\par In this paper, we will study the time evolution of the charge imbalance resolved negativity after a local joining quench for both adjacent and disjoint intervals and explain our results in terms of quasi-particle picture. The basic idea of quasi-particle picture is that quasi-particles are produced at the joining point. These quasi-particles may be viewed as entangled pairs which carry entanglement information. The entanglement negativity can be built immediately when the entangled pairs arrive at two intervals separately.
\par The plan of this paper is as follows. In section \ref{section2}, we briefly introduce the CFT setup for local joining quenches. In section \ref{s3}, we briefly review the concept of symmetry resolved entanglement and negativity. In section \ref{section3}, we review how to compute the time evolution of charged logarithmic negativity and charged imbalance resolved negativity after a local joining quench.  In section \ref{section4}, we consider the time evolution of charged logarithmic negativity between two adjacent and disjoint intervals in the local joining quench protocol. In section \ref{section5}, we will calculate the charge imbalance resolved negativity from the results obtained in the previous sections. In section \ref{section6}, we compare our CFT predictions with the numerical results in the complex harmonic chain. We further explain our results for the charged logarithmic negativity based on the quasi-particle picture. Finally,
in section \ref{section7}, we conclude our results and discuss some interesting problems which are worth investigating in the future.
\section{Local joining quench}\label{section2}
\par In this paper, we will consider the $1+1$ dimensional complex free scalar field theory
with the Euclidean action given by
\be
\mathcal{A}=\int d^2x\partial_{\mu}\phi^{\dg}\partial_{\mu}\phi.
\ee
This action exhibit a $U(1)$ symmetry, i.e. the phase transformation of the field $\phi\rightarrow e^{i\theta}\phi, \phi^{\dg}\rightarrow e^{-i\theta}\phi^{\dg}$ leaves the action invariant. Initially, we have two independent systems with one spatial dimension and prepare them in their own ground states. At $t=0$, we join the two systems at their endpoints together.
\par Now, we briefly review the CFT approach to the local joining quench following Ref.\cite{Wen:2015qwa}. We first introduce the time evolution of the density matrix $\rho(t)=|\phi(x,t)\rangle\langle \phi(x,t)|$ with $|\phi(x,t)\rangle=e^{-\mathrm{i}Ht}|\phi_0(x)\rangle$. Then
\be
\langle \phi^{\prime\prime}(x^{\prime\prime})|\rho(t)|\phi'(x')\rangle
=Z^{-1}\langle \phi^{\prime\prime}(x^{\prime\prime})|e^{-\mathrm{i}Ht-\epsilon H}|\phi_0(x)\rangle\langle\phi_0(x)|e^{+\mathrm{i}Ht-\epsilon H}|\phi'(x')\rangle,
\ee
where two factors $e^{-\epsilon H}$ are used to make the path integral representation of this element absolutely convergent, and
$Z=\langle \phi_0| e^{-2\epsilon H} | \phi_0\rangle $ is the normalization factor.
\begin{figure}
\centering
\includegraphics[scale=0.8]{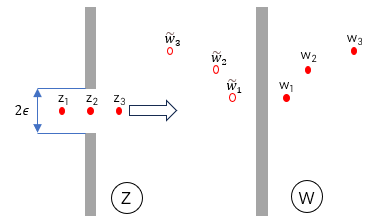}
\caption{The conformal mapping in Eq.(\ref{conformalMap}), based on which the
$z$-plane is mapped to a right half plane (RHP) with $\text{Re}\ (w)>0$.}\label{ljqsetup}
\end{figure}
We can express the density matrix using the path integral on a modified world-sheet, as shown in Fig.~\ref{ljqsetup}, where the physical cut corresponds to having a slit parallel to the imaginary time axis, giving two parts with one starting from $-\infty$ up to $\tau_1=-\epsilon-\mathrm{i}t$ and the other starting from $\tau_2=+\epsilon-\mathrm{i}t$ to $+\infty$ in the complex $z$-plane. For computation simplicity, we map the $z$-plane to a right half plane (RHP), with $\Re (w)>0$ in terms of the conformal mapping
\begin{equation}
\label{conformalMap}
w(z)=\frac{z}{\epsilon}+\sqrt{\left(\frac{z}{\epsilon}\right)^2+1},
\quad  z(w)=\epsilon \frac{w^2-1}{2w}.
\end{equation}
\section{Symmetry resolved entanglement}\label{s3}
In this section, we will briefly review the decomposition of entanglement entropy and negativity under a global internal symmetry.
\subsection{Symmetry resolved entropies}
\par Let us first review the basic idea about the symmetry resolved entanglement entropy. If our system exhibit a global $U(1)$ symmetry generated by a local charge $Q$ which commute with the density matrix, i.e. $[\rho, Q]=0$, then we have $[\rho_A,Q_A]=0$. Thus $\rho_A$ admits charge decomposition according to eigenvalues $q$ of the local charge $Q_A$
\be
\rho_A=\oplus_q\mathcal{P}_q\rho_A=\oplus_qp(q)\rho_A(q),\quad p(q)=\Tr(\mathcal{P}_q\rho_A),
\ee
where $\mathcal{P}_q$ is the projection operator which projects the space to the eigenspace corresponding to eigenvalue $q$. The projector $\mathcal{P}_q$ has the following integral represent
\be
\label{projectors}
\mathcal{P}_{q}=\int_{0}^{2\pi}\frac{d\m}{2\pi}e^{-\mathrm{i}\m q}e^{\mathrm{i}\m Q_A}.
\ee
The symmetry resolved R\'enyi entropies are defined as
\be
S_n(q)=\frac{1}{1-n}\log\Tr[\rho_A(q)]^n.
\ee
Symmetry resolved entanglement entropy can be obtained by taking the replica limit $S(q)=\lim_{n\rightarrow 1}S_n(q)$.
\par Regarding the integral represent of the project operator, it's convenient to first introduce the charged moments of $\rho_A$,
\be\label{Znmiu}
Z_n(\m)=\Tr(e^{\mathrm{i}\m Q_A}\rho_A^n).
\ee
Then it's sufficient to compute its Fourier transform
\be
Z_n(q)=\int_0^{2\pi}\frac{d\m}{2\pi}e^{-\mathrm{i}q\m}Z_n(\m)
\ee
to obtain the R\'enyi entropies of the sector with charge $q$ as
\be
S_n(q)=\frac{1}{1-n}\log\left[\frac{Z_n(q)}{Z_1(q)^n}\right].
\ee
\subsection{Charge imbalance resolved negativity}
\par Now we turn to the symmetry decomposition of entanglement negativity. We briefly recall some results about the charge imbalance resolved negativity under the $U(1)$ symmetry generated by a local charge $Q$ (see \cite{Cornfeld:2018wbg,Murciano:2021djk,Gaur:2022sjf,Parez:2022xur,Chen:2022gyy} for more details). We consider the case of
a bi-partition in two complementary subsystems $A$ and $B$, in which the subsystem $A$ is partitioned into two complementary subsystems $A_1$ and $A_2$, we denote the corresponding charge operators as $Q_1$ and $Q_2$. Because of locality, we can get $Q=Q_A+Q_B$ and $Q_A=Q_1+Q_2$.
From the relation $[\rho_A,Q_A]=0$, performing the partial transpose with respect to the degrees of freedom of subsystem $A_2$, we have
\be
\label{cr}
[\rho_A^{T_2},\mQ_A]=0,
\quad \mQ_A=Q_1-Q_2^{T_2},
\ee
where we define the operator $\mQ_A$ as charge imbalance operator. We denote the eigenvalue of the charge imbalance operator $\mQ_A$ by $\mathrm{q}$, which is an integer. We also denote the projectors on the corresponding eigenspace by $\mathcal{P}_{\mathrm{q}}$.
Then $\rho_A^{T_2}$ has a form of block matrix, each block is characterized by different eigenvalues ${\mathrm{q}}$, thus we can write
\be
\label{rhot2q}
\rho_A^{T_2}({\mathrm{q}})=\frac{\mP_{\mathrm{q}} \rho_A^{T_2}}{\Tr (\mP_{\mathrm{q}} \rho_A^{T_2})}.
\ee
It is enough to write
\begin{equation}
    \rho_A^{T_2}=\bigoplus_{{\mathrm{q}\in\mathbb{Z}}}p({\mathrm{q}})\rho_A^{T_2}({\mathrm{q}}),
\end{equation}
where $p({\mathrm{q}})=\Tr (\mP_{\mathrm{q}} \rho_A^{T_2})$ is the probability of finding ${\mathrm{q}}$ as the outcome of a measurement of $\mQ_A$.
The charge imbalance resolved  negativity is defined as
\begin{equation}
\label{totalne}
    \mN(\mathrm{q})=\frac{\mathrm{Tr}|\rho_A^{T_2}(\mathrm{q})|-1}{2}.
\end{equation}
The total negativity obtained by the sum of charge imbalance resolved negativity weighted by the corresponding probability
\be
\mN=\sum_{\mathrm{q}}p(\mathrm{q})\mN(\mathrm{q}).
\ee
It is useful to define the charge imbalance resolved R\'enyi negativity as
\be
\mathcal{N}_n(\mathrm{q}) = \frac{N_n(\mathrm{q})-1}{2},\quad \text{with}\quad
N_n(\mathrm{q})=\Tr[(\rho_A^{T_2}({\mathrm{q}}))^n]=\frac{\Tr[\mathcal{P}_{\mathrm{q}}(\rho_A^{T_2})^n]}{p({\mathrm{q}})^n}.
\ee
Then the charged imbalance resolved negativity is obtained by taking limit
\be
\mathcal{N}(\mathrm{q})= \lim_{n_e \to 1} \mathcal{N}_{n_e}(\mathrm{q}).
\ee
\par Based on the integral representation of the projector $\mathcal{P}_{\mathrm{q}}$, one finds that it's useful to define the charged R\'enyi negativity and its Fourier transform
\be\label{Rnmu}
N_n(\m)=\Tr[(\rho_A^{T_2})^ne^{\mathrm{i}\m\mathcal{Q}_A}],
\quad \mathcal{Z}_{T_2,n}(\mathrm{q})=\int_0^{2\pi}\frac{d\m}{2\pi} e^{-\mathrm{i}\m\mathrm{q}}N_n(\m).
\ee
The charged R\'enyi logarithmic negativity is simply given by
\be
\mE_n(\m)=\log N_{n}(\m).
\ee
The charged negativity is obtained by taking the replica limit $n_e\rightarrow 1$ in $N_{n_e}(\m)$
\be
\label{cn}
N(\m)=\lim_{n_e \to 1} N_{n_e}(\m),\qquad \mE(\m)= \log N(\m).
\ee
The probability distribution can be obtained as
\be
\label{p(q)}
p(\mathrm{q})=\int_0^{2\pi}\frac{d\m}{2\pi} e^{-\mathrm{i}\m {\mathrm{q}}}N_1(\m)
=\int_0^{2\pi}\frac{d\m}{2\pi} e^{-\mathrm{i}\m {\mathrm{q}}} e^{\mE_1(\m)}.
\ee
Then it is easy to derive
\be
N_n({\mathrm{q}})=\frac{\mathcal{Z}_{T_2,n}(\mathrm{q})}{p(\mathrm{q})^n}.
\ee
We also introduce the quantity
\be
\label{zt2}
\mathcal{Z}_{T_2}(\mathrm{q})\equiv\lim_{n_e\rightarrow 1}\mathcal{Z}_{T_2,n_e}(\mathrm{q})
=\int_0^{2\pi}\frac{d\m}{2\pi} e^{-\mathrm{i}\m\mathrm{q}}e^{\mE(\m)},
\ee
from which we can obtain the charge imbalance resolved negativity as
\be
\label{cirn}
\mathcal{N}(\mathrm{q})=\frac12\left(\frac{\mathcal{Z}_{T_2}(\mathrm{q})}{p(\mathrm{q})}-1\right)=\frac12\left(\frac{\int_0^{2\pi}\frac{d\m}{2\pi} e^{-\mathrm{i}\m\mathrm{q}}e^{\mE(\m)}}{\int_0^{2\pi}\frac{d\m}{2\pi} e^{-\mathrm{i}\m {\mathrm{q}}} e^{\mE_1(\m)}}-1\right).
\ee
\section{Charged R\'enyi negativity and fluxed twist field}
\label{section3}
\par
\begin{figure}
\centering
\includegraphics[scale=1]{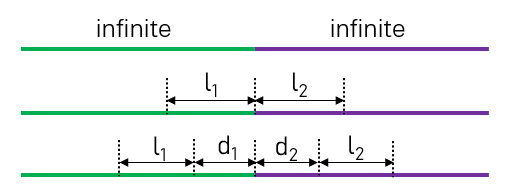}
\caption{Different cases we consider here: semi-infinite intervals (up), two adjacent intervals (middle) and two disjoint intervals (bottom). }\label{di}
\end{figure}

 As shown in Fig.~\ref{di}, we consider the cases of two semi-infinite intervals (up), two adjacent intervals (middle), and two disjoint intervals (bottom). In these three cases, we will respectively calculate the dynamics of charged logarithmic negativity after a local joining quench in CFT. In this section, we will briefly review the method of computing the charged logarithmic negativity using the fluxed twist field \cite{Cornfeld:2018wbg}.
\par The connection between the charged negativity and the fluxed twist fields was first derived in reference \cite{Cornfeld:2018wbg}. We will review some basic facts about the charged logarithmic negativity in the $(1+1)$ dimensional CFT (see \cite{Chen:2021nma} for more details). We assume that our system exhibits a global $U(1)$ symmetry and the subsystem $A$ consists of $N$ disjoint intervals.
\par We can view the charged moments of $\rho_A$ i.e. $Z_n(\m)$ (cf. eq.~(\ref{Znmiu})) as the partition function on the Riemann surface $\mR_{n,N}$
pierced by an Aharonov-Bohm flux, such that the total phase accumulated by the field upon going through the entire surface is $\m$. We can define the operator $\mV_{\m}$ and the local twist field $\mT_n$ which generate the total phase $\m$ and the $n$-sheet Riemann surface $\mR_{n,N}$, respectively. We call the fusion of $\mV_{\m}$ and $\mT_n$ as fluxed twist field denoted by $\mT_{n,\m}$ \cite{Goldstein:2017bua}. Then the partition function on the fluxed Riemann surface $\mR^{(\m)}_{n,N}$  is obtained from the $2N$-point function of these fluxed twist operators. By using the conformal mapping defined in Eq.~(\ref{conformalMap}), the problem of local joining quench is reduced to the computation of the correlation functions of the fluxed twist fields in the RHP. In free complex boson CFT, the conformal dimension of the fluxed twist field and the fluxed anti-twist field are the same, it reads \cite{Chen:2021nma}
\be
\D_{n,\m}=\frac{1}{6}\lt(n-\frac{1}{n}\rt)-\frac{\m^2}{4\pi^2n}+\frac{\m}{2\pi n}.
\ee
\par For simplicity, in the following part, we will focus on the case $N=2$, i.e. we will consider the case where the subsystem $A$ consists of two disjoint intervals: $A_1=[l_1,l_2]$ ,$A_2=[l_3,l_4]$ and $A=A_1\cup A_2$.
\par For two adjacent intervals, the charged moments of the partially transposed RDM or the charged R\'enyi negativity equivalently has the following structure
\begin{equation}
\label{threepoint}
\begin{aligned}
\operatorname{Tr}[\left(\rho_A^{T_2}\right)^n  e^{\mathrm{i}\m\mQ_A} ]& =\left\langle\mathcal{T}_{n,\mu}\left(z_1\right)
\tilde{\mathcal{T}}_{n,\mu}^2\left(z_2\right) \mathcal{T}_{n,\mu}\left(z_3\right)\right\rangle
\end{aligned},
\quad z_i=l_i+i\tau.
\end{equation}
 By using the conformal mapping defined in Eq.~(\ref{conformalMap}), the three-point function on the $z$-plane is mapped onto the RHP, we have
\begin{equation}
\label{threepoint}
\begin{aligned}
\operatorname{Tr}[\left(\rho_A^{T_2}\right)^n  e^{\mathrm{i}\m\mQ_A} ]
& =\prod_{i=1}^3\left|\frac{d w}{d z}\right|_{z_i}^{\Delta_{(i)}}\left\langle\mathcal{T}_{n,\mu}\left(w_1\right)  \tilde{\mathcal{T}}_{n,\mu}^2\left(w_2\right) \mathcal{T}_{n,\mu}\left(w_3\right)\right\rangle_{\mathrm{RHP}},
\end{aligned}
\end{equation}
where the scaling dimensions $\Delta_{(1)}=\Delta_{(3)}=\Delta_{n,\m}$ and $\Delta_{(2)}=\Delta_{n,\m}^{(2)}$. The scaling dimension $\Delta^{(2)}_{n,\m}$ of the operator $\mathcal{T}^2_{n,\m}$ reads \cite{Chen:2022gyy}
\be
\D_{n,\m}^{(2)}=
\begin{cases}
\D_{n,2\m},\quad \text{odd}~n\\
2\D_{\frac{n}{2},\m},\quad \text{even}~n
\end{cases}
\ee
The three-point function on the RHP has the standard form
\be
\label{3pf}
\begin{split}
\langle\mathcal{T}_{n,\m}(w_{1})\tilde{\mathcal{T}}^2_{n,\m}(w_{2})\mathcal{T}_{n,\m}(w_{3})\rangle_{\text{RHP}}
=&\frac{c_{n,\m}}{\prod_{i=1}^3|(w_i-\widetilde{w}_i)/a|^{\Delta_{(i)}}}\left(
\frac{\eta_{1,3}^{\Delta_{n,\mu}^{(2)}-2\Delta_{n,\mu}}}{\eta_{1,2}^{\Delta_{n,\mu}^{(2)}}\eta_{2,3}^{\Delta_{n,\mu}^{(2)}}}
\right)^{1/2} \mathcal{F}_{n}(\{\eta_{j,k}\}).
\end{split}
\ee
where the constant $c_{n,\m}$ known for some specific theories are nonuniversal and  $a$ is an UV cutoff. The cross ratios $\eta_{i,j}=\frac{(w_i-w_j)(\widetilde{w}_i-\widetilde{w}_j)}{(w_i-\widetilde{w}_j)(\widetilde{w}_i-w_j)}$ with $\widetilde{w}_i=-\bar{w}_i$, which is the image of $w_i$ as shown in Fig.\ref{ljqsetup}. The nonuniversal function $\mathcal{F}_{n}(\{\eta_{j,k}\})$ depends on the full operator content of the theory and is very difficult to calculate.
\par For two disjoint intervals, the charged moments of the partially transposed RDM after doing a conformal map onto the RHP, we have
\be\label{4ptf}
\begin{aligned}
\Tr[(\rho_A^{T_2})^ne^{\mathrm{i}\m\mQ_A}]&
=\langle\mathcal{T}_{n,\m}(z_1)\tilde{\mathcal{T}}_{n,\m}(z_{2})\tilde{\mathcal{T}}_{n,\m}(z_3)\mathcal{T}_{n,\m}(z_{4})\rangle\\
&=\prod_{i=1}^4\left|\frac{d w}{d z}\right|_{z_i}^{\Delta_{(i)}}
\left\langle\mathcal{T}_{n,\m}\left(w_1\right) \tilde{\mathcal{T}}_{n,\m}\left(w_2\right) \tilde{\mathcal{T}}_{n,\m}\left(w_3\right) \mathcal{T}_{n,\m}\left(w_4\right)\right\rangle_{\mathrm{RHP}}\\
\end{aligned}
\ee
where the scaling dimensions $\Delta_{(1)}=\Delta_{(2)}=\Delta_{(3)}=\Delta_{(4)}=\Delta_{n,\m}$.
Here, comparing to the original charged moments of the RDM, we have exchanged the fluxed twist field $\mathcal{T}_{n,\m}$ and $\tilde{\mathcal{T}}_{n,\m}$ at the endpoints of $A_2$ while keep the others unchanged.
By global conformal symmetry, the four-point function on the RHP has the following form
\begin{equation}
\label{4pf}
\begin{split}
& \left\langle\mathcal{T}_{n,\m}\left(w_1\right) \tilde{\mathcal{T}}_{n,\m}\left(w_2\right) \tilde{\mathcal{T}}_{n,\m}\left(w_3\right) \mathcal{T}_{n,\m}\left(w_4\right)\right\rangle_{\mathrm{RHP}}\\
=&\frac{c_{n,\m}^2}{\prod_{i=1}^4|(w_i-\widetilde{w}_i)/a|^{\Delta_{n,\mu}}}
\frac{1}{\eta_{1,2}^{\Delta_{n,\mu}}\eta_{3,4}^{\Delta_{n,\mu}}}
\left(\frac{\eta_{1,4}\eta_{2,3}}{\eta_{1,3}\eta_{2,4}}\right)^{\Delta_{n,\mu}^{(2)}/2-\Delta_{n,\mu}} \mathcal{F}_{n}(\{\eta_{j,k}\}),
\end{split}
\end{equation}
where the $\Delta_{n,\m}^{(2)}$ appears since that the  Eq.~(\ref{3pf}) should be reproduced when we take the adjacent limit $z_3 \rightarrow z_2$ in Eq.~(\ref{4pf}).
\section{Charged logarithmic negativity after a local joining quench}
\label{section4}
\subsection{Two adjacent intervals}
\subsubsection{Semi-infinite intervals}
\par As shown in Fig.\ref{di}, it is convenient to consider the simplest case, in which the total system is made of two semi-infinite parts $A_1$ and $A_2$. In this case, the time evolution of charged R\'enyi negativity
is governed by $\left\langle \mathcal{T}^2_{n,\m}(z_1)\right\rangle$ on the strip where $\mathcal{T}^2_{n,\m}(z_1)$ is a single fluxed twist field inserted at
$ z_1=l+\mathrm{i}\tau$ in the z-plane. Here we choose $l=0$, we get $ z_1=\mathrm{i}\tau$. The expectation value of $\mathcal{T}^2_{n,\m}(z_1)$ has the standard form
\begin{equation}
\left\langle\mathcal{T}^2 _{n,\mu} \left(z_1\right)\right\rangle={c}_{n,\mu}
\left(\left|\frac{d w}{d z}\right|_{z_1} \frac{a}{2 \operatorname{Re}\left(w_1\right)}\right)^{\Delta_{n,\mu}^{(2)}}.
\end{equation}
After the conformal mapping defined in Eq.~(\ref{conformalMap}), we obtain
\begin{equation}
w_1=\mathrm{i}\frac{\tau}{\epsilon}+\frac{1}{\epsilon}\sqrt{\epsilon^2-\tau^2},
\end{equation}
then we can get
\begin{equation}
\left|\frac{dw}{dz}\right|_{z_1}=\frac{1}{\sqrt{\epsilon^2-\tau^2}}.
\end{equation}
Considering $\tau$ as a real number, and only at the end of the computation we analytically continue $\tau\to \mathrm{i}t$. Finally, we find
\begin{equation}
\left\langle \mathcal{T}^2_{n,\mu}(z_1)\right\rangle=c_{n,\mu}
\left(\frac{a\epsilon}{2(\epsilon^2+t^2)}\right)^{\Delta^{(2)}_{n,\mu}}.
\end{equation}
Now it is straightforward to derive the charged R\'enyi logarithmic negativity
\be
\mathcal{E}_n(\mu)=\Delta_{n, \mu}^{(2)} \ln \frac{a \epsilon}{2\left(\epsilon^2+t^2\right)}+\ln c_{n, \mu}.
\ee
The charged logarithmic negativity is obtained by taking replica limit $n_e \rightarrow$ 1
\be
\mathcal{E}(\mu)=h(\mu) \ln \frac{a \epsilon}{2\left(\epsilon^2+t^2\right)}+\tilde{c}_{\m},
\qquad h(\mu)=-\frac12-\frac{\m^2}{\pi^2}+\frac{2\m}{\pi},
\ee
where $\tilde{c}_{\m}=\lim_{n_e\rightarrow 1} \ln c_{n_e, \mu}$. At $t=0$, requiring
\be
\mathcal{E}(\mu) \big|_{t=0}=h(\mu) \ln \frac{a}{2\epsilon}+\tilde{c}_{\m}=0,
\ee
it is enough to fix the factor $\tilde{c}_{\m}$. We get
\be
\label{semiin}
\mathcal{E}(\mu)=h(\mu) \ln \frac{\epsilon^2}{\epsilon^2+t^2}.
\ee
In the limit $t\gg\epsilon$, we find
\be
\mathcal{E}(\mu)=-2 h(\mu) \ln \frac{t}{\epsilon}.
\ee
When $\m=0$, we have
\be
\mathcal{E}(\mu=0)=\ln \frac{t}{\epsilon},
\ee
which agrees with the result obtained in Ref.\cite{Wen:2015qwa}.
\par Particularly, for $n=1$, we have
\be
\mathcal{E}_1(\mu)
=h_1 (2\mu) \ln \frac{\epsilon^2}{\epsilon^2+t^2},\quad
h_1 (\mu)=-\frac{\m^2}{ 4 \pi^2}+\frac{\m}{ 2\pi}.
\ee
\subsubsection{Symmetric finite intervals}
\par In this part, we consider the case of symmetric finite intervals with $l_1=l_2=l$, namely, $A_1\in[-l,0]$ and $A_2\in(0,l]$ (cf. Fig.~\ref{di}). In this case, by studying the three-point function in Eq.~(\ref{threepoint}), quench dynamics of the charged R\'enyi negativity between two adjacent intervals can be obtained. $\mathcal{F}(\{\eta_{j,k}\})$ is just a constant in the limits $\eta_{i,j}\to0$, $1$, or $\infty$. For symmetric intervals, it has been found that one always has $\eta_{ij}=1$ or $0$ for the cases $t\ll l$, $t=l+0^-$ and $t>l$ \cite{Wen:2015qwa}. Thus our results are universal for the above three cases.
\par Neglecting various non-universal terms and using Eqs. (\ref{3pf}) and (\ref{cn}), we obtain
\be
\mathcal{E}(\mu)=
\begin{cases}
h(\mu) \ln \frac{\epsilon^2}{\epsilon^2 +t^2}+h(\m)\ln \frac{l+t}{l-t}+2 h_1(\mu) \ln \frac{1}{2 l}, & t<l \\
h(\mu) \ln \frac{\epsilon^2}{\epsilon^2 +t^2}+h(\m)\ln \frac{4 t^2}{\epsilon l}+
h_1(\m)\ln \frac{(t-l)^2}{l^2  [4  (t-l)^2 +4t^2 \epsilon^2]}
, & t>l
\end{cases}
\ee
In the space-time scaling limit $l,t\gg\epsilon$, by subtracting the charged logarithmic negativity at $t=0$ , we end up with
\be
\label{taisfimE}
\mathcal{E}(\mu)=
\begin{cases}
h(\mu) \ln \frac{\epsilon^2 (l+t)}{\left( \epsilon ^2+t ^2 \right)(l-t)}, & t<l \\
h(\mu) \ln \frac{4 \epsilon}{ l}, & t>l
\end{cases}
\ee
In the regime $t\ll l$, we get
\be
\mathcal{E}(\mu)=h(\mu) \ln \frac{\epsilon^2}{\epsilon^2+t^2},
\ee
which is the same as Eq.~(\ref{semiin}). For $t\le l$, our results are not accurate, because we neglect the non-universal functions $\mathcal{F}(\{\eta_{j,k}\})$ which may not be constant. Thus we may not obtain the correct scaling behavior.  For $t>l$, we can compute the ground-state value of $\mathcal{E}(\mu)$ as
\be
\mathcal{E}_G(\m)=-h(\mu) \ln \frac{ l}{4 \epsilon},
\ee
which is consistent with the result in Ref.~\cite{Wen:2015qwa}.
\subsubsection{Asymmetric finite intervals}
\par We consider the case of asymmetric finite intervals with $A_1\in[-l_1,0]$ and $A_2\in(0,l_2]$. Without loss of generality, we assume $l_1<l_2$. It has been found that one always has $\eta_{ij}=1$ or $0$ for the cases $t\ll l_1$, $t=l_2+0^-$ and $t>l_2$\cite{Wen:2015qwa}. Our results are universal in these cases. By neglecting the non-universal terms, the charged logarithmic negativity can be written as
\be
\mathcal{E}(\mu)=\frac{h(\m)}{2}\ln{\left(\frac{\eta_{1,3}}{\eta_{1,2}\eta_{2,3}}\right)}.
\ee
In the limit $l,t\gg\epsilon$, we have
\be
\mathcal{E}(\mu)=
\begin{cases}
\frac{h(\mu)}{2}  \ln \frac{\epsilon^4 (l_1+t)(l_2+t)}{t ^4(l_1-t) (l_2-t)} , & t<l_1\\
\frac{h(\mu)}{2}  \ln \frac{4 \epsilon^2 (l_1+l_2)(t^2-l_1^2)}{(l_2-l_1)l_1^2t^2}, &l_1< t<l_2\\
h(\mu)\ln \frac{2 \epsilon(l_1+l_2)}{l_1l_2}, &t>l_2
\end{cases}
\ee
If we take $l_1=l_2$, the result in Eq.~(\ref{taisfimE}) is reproduced. For $t\le l_2$, similar to the symmetric case, our results are not accurate, because we neglected the nonuniversal functions $\mathcal{F}(\{\eta_{j,k}\})$ which may not be constant. Thus one has to calculate $\mathcal{F}(\{\eta_{j,k}\})$ for different CFTs. For $t>l_2$, we obtain the ground-state value of the charged logarithmic negativity as
\be
\mathcal{E}_G(\m)=-h(\mu)\ln \frac{l_1l_2}{2 \epsilon(l_1+l_2)},
\ee
which is consistent with the result in Ref.~\cite{Wen:2015qwa}.
\subsection{Two disjoint intervals}
\subsubsection{Symmetric finite intervals}
In this part, we consider the case of symmetric finite intervals with $A_1\in[-d-l,-d]$ and $A_2\in[d,d+l]$ (cf. Fig.~\ref{di}). In this case, by studying the four-point function in Eq.~(\ref{4ptf}), quench dynamics of the charged R\'enyi negativity between two disjoint symmetric intervals can be obtained. As explicitly calculated in Ref. \cite{Asplund:2013zba}, $\mathcal{F}(\{\eta_{j,k}\})$ is simply a constant in the limit $l/d \ll 1$. The calculations are similar to the two adjacent symmetric finite intervals case. In the limit $l,t\gg\epsilon$, by neglecting various non-universal terms, using Eqs. (\ref{4pf}) and (\ref{cn}) and subtracting the charged logarithmic negativity at $t=0$, we have
\be
\label{tdisfimE}
\mathcal{E}(\mu)=
\left\{
\begin{aligned}
&0   &t<d\\
& h(\mu)\ln\frac{\epsilon dl(d+l+t)}{(2d+l)(d+l-t)(t^2-d^2)} &d<t<d+l\\
&h(\mu)\ln\frac{4 d(d+l)}{(2d+l)^2}  &t>d+l
\end{aligned}
\right.
\ee
\subsubsection{Asymmetric finite intervals}
In this case, we consider the case of asymmetric finite intervals with $A_1\in[-d_1-l_1,d_1]$ and $A_2\in[d_2,d_2+l_2]$. Without loss of generality, we assume $l_1=l_2=l$ and $d_1<d_2\leq d_1+l$. By neglecting the nonuniversal terms, the charged logarithmic negativity can be written as
\be
\mathcal{E}(\mu)=\frac{h(\m)}{2}\ln{\left(\frac{\eta_{1,4}\eta_{2,3}}{\eta_{1,3}\eta_{2,4}}\right)}.
\ee
In the limit $l,t\gg\epsilon$, we have
\be
\mathcal{E}(\m)=
\left\{
\begin{aligned}
&0&t<d_1\\
&\frac{h(\mu)}{2}\ln\frac{(d_1+d_2)(d_2-t)(d_2+l+t)(d_2-d_1+l)}{(d_1+d_2+l)(d_2+l-t)(d_2+t)(d_2-d_1)}&d_1<t<d_2\\
& \frac{h(\mu)}{2} \ln\frac{\epsilon^2(d_1+d_2)^2(d_1+l-d_2)(d_2+l-d_1)(d_1+l+t)(d_2+l+t)}{4(d_1+d_2+l)^2(d_1+l-t)(d_2+l-t)(t^2-d_1^2)(t^2-d_2^2)}&d_2<t<d_1+l\\
&\frac{h(\mu)}{2}\ln\frac{[t^2-(d_1+l)^2](d_1+d_2)^2(d_2+l-d_1)(d_1+d_2+2l)}{(d_2-d_1)(d_1+d_2+l)^3(t^2-d_1^2)}&d_1+l<t<d_2+l\\
&h(\mu)\ln\frac{(d_1+d_2)(d_1+d_2+2l)} {(d_1+d_2+l)^2}&t>d_2+l\\
\end{aligned}
\right.
\ee
When taking $d_1=d_2=d$, our result in Eq.~(\ref{tdisfimE}) can be reproduced.
\section{Charge imbalance resolved negativity after a local joining quench}
\label{section5}
\par In this section, we consider the symmetric cases. For simplicity, we introduce some notations
\be
\begin{split}
&f_1 \equiv \ln \frac{\epsilon^2}{\epsilon^2+t^2}. \\
&f_2 \equiv  \ln \frac{\epsilon^2 (l+t)}{\left( \epsilon ^2+t ^2 \right)(l-t)}.\\
&f_3 \equiv  \ln \frac{4 \epsilon}{ l}.\\
&f_4 \equiv \ln\frac{\epsilon dl(d+l+t)}{(2d+l)(d+l-t)(t^2-d^2)}.\\
&f_5 \equiv \ln\frac{4 d(d+l)}{(2d+l)^2}.
\end{split}
\ee
\subsection{Two adjacent intervals}
\subsubsection{Semi-infinite intervals}
During this region, we have
\be
\mathcal{E}(\mu)=f_1 h(\mu),\quad
\mathcal{E}_1(\mu)=f_1 h_1(2\mu) ,
\ee
when $\m=0$, we obtain
\be
\label{semie0}
\mathcal{E}(0)=f_1 h(0).
\ee
By using Eq.~(\ref{p(q)}), after normalization, we find that the probability distribution is given by
\be
\label{semiq}
p(\mathrm{q})=
-\frac{1+(-1)^\mathrm{q} }{\pi^2 \mathrm{q}^2/f_1 +f_1 }.
\ee
After Fourier transformation in Eq.~(\ref{zt2}), we have
\be
\mathcal{Z}_{T_2}(\mathrm{q})=\frac{- 2 f_1}{e^{\frac{f_1}{2}}(\pi^2 \mathrm{q}^2 +4f_1^2)}.
\ee
Thus the charge imbalance resolved negativity is obtained from Eq.~(\ref{cirn}) as
\be
\label{semicirn}
\mathcal{N}(\mathrm{q})=\frac{\pi^2 \mathrm{q}^2 +f_1^2}{e^{\frac{f_1}{2}}(\pi^2 \mathrm{q}^2 +4f_1^2)(1+(-1)^\mathrm{q})}-\frac{1}{2}.
\ee
\subsubsection{Symmetric finite intervals}
$\pmb{t<l}$\\
\\
In this region, we have
\be
\mathcal{E}(\mu)=f_2 h(\mu) ,\quad
\mathcal{E}_1(\mu)=f_2 h_1(2\mu) .
\ee
The probability distribution is
\be
p(\mathrm{q})=
-\frac{1+(-1)^\mathrm{q} }{\pi^2 \mathrm{q}^2/f_2 +f_2 }.
\ee
After Fourier transformation, we find
\be
\mathcal{Z}_{T_2}(\mathrm{q})=\frac{- 2 f_2}{e^{\frac{f_2}{2}}(\pi^2 \mathrm{q}^2 +4f_2^2)}.
\ee
Thus the charge imbalance resolved negativity is
\be
\mathcal{N}(\mathrm{q})=\frac{\pi^2 \mathrm{q}^2 +f_2^2}{e^{\frac{f_2}{2}}(\pi^2 \mathrm{q}^2 +4f_2^2)(1+(-1)^\mathrm{q})}-\frac{1}{2}.
\ee
$\pmb{t>l}$\\
\\
In this region, we have
\be
\mathcal{E}(\mu)=f_3 h(\mu)  ,\quad
\mathcal{E}_1(\mu)=f_3 h_1(2\mu) .
\ee
The probability distribution can be computed similarly, and the final result is
\be
p(\mathrm{q})=
-\frac{1+(-1)^\mathrm{q} }{\pi^2 \mathrm{q}^2/f_3 +f_3 }.
\ee
After Fourier transformation, we find
\be
\mathcal{Z}_{T_2}(\mathrm{q})=\frac{-  2 f_3}{e^{\frac{f_3}{2}}(\pi^2 \mathrm{q}^2 +4f_3^2)}.
\ee
Then the charge imbalance resolved negativity is
\be
\mathcal{N}(\mathrm{q})=
\frac{\pi^2 \mathrm{q}^2 +f_3^2}{e^{\frac{f_3}{2}}(\pi^2 \mathrm{q}^2 +4f_3^2)(1+(-1)^\mathrm{q})}-\frac{1}{2}.
\ee
\subsection{Two disjoint symmetric finite intervals}
$\pmb{t<d}$\\
\\
At this early time stage, we have $\mathcal{E}(\m)=\mathcal{E}_1(\m)$, therefore
\be
\mathcal{N}(\mathrm{q})=0.
\ee
$\pmb{d<t<d+l}$\\
\\
In this time region, we have
\be
p(\mathrm{q})=
-\frac{1+(-1)^\mathrm{q} }{\pi^2 \mathrm{q}^2/f_4 +f_4 },
\ee
\be
\mathcal{Z}_{T_2}(\mathrm{q})=\frac{-2  f_4}{e^{\frac{f_4}{2}}(\pi^2 \mathrm{q}^2 +4f_4^2)} .
\ee
The charge imbalance resolved negativity is given by
\be
\mathcal{N}(\mathrm{q})=\frac{\pi^2 \mathrm{q}^2 +f_4^2}{e^{\frac{f_4}{2}}(\pi^2 \mathrm{q}^2 +4f_4^2)(1+(-1)^\mathrm{q})}-\frac{1}{2}.
\ee
$\pmb{t>d+l}$\\
\\
In this time region, we have
\be
p(\mathrm{q})=
-\frac{1+(-1)^\mathrm{q} }{\pi^2 \mathrm{q}^2/f_5 +f_5 },
\ee
\be
\mathcal{Z}_{T_2}(\mathrm{q})=\frac{- 2 f_5}
{e^{\frac{f_5}{2}}(\pi^2 \mathrm{q}^2 +4f_5^2)} .
\ee
The charge imbalance resolved negativity is
\be
\mathcal{N}(\mathrm{q})=\frac{\pi^2 \mathrm{q}^2 +f_5^2}{e^{\frac{f_5}{2}}(\pi^2 \mathrm{q}^2 +4f_5^2)(1+(-1)^\mathrm{q})}-\frac{1}{2}.
\ee
\par From the explicit calculation in this section, it's clear that the charge imbalance resolved negativity are all $\mathrm{q}$ dependent. The equipartition property \cite{Xavier:2018kqb, Bonsignori:2019naz} observed in the symmetry resolved entanglement entropy breaks down explicitly in our case.
\subsection{Total negativity}
\par To check our results, we should recover the total negativity from the charge imbalance resolved negativity from the formula
\be
\sum_{\mathrm{q}=-\inf}^{+\inf} p(\mathrm{q}) \mN(\mathrm{q})=\frac{e^{-\frac{f}{2}}-1}{2}=\frac{e^{\mE(\m=0)}-1}{2}=\mN,
\ee
where $\mE(\m=0)=\mE$ is the total logarithmic negativity defined in Eq.~(\ref{logNeg}).
Here we have used the following formula
\be
\label{eq:pqSums}
\begin{split}
\sum_{\mathrm{q}=-\inf}^{+\inf} p(\mathrm{q}) =- \sum_{\mathrm{q}=-\inf}^{+\inf} \frac{f}{\pi^2 \mathrm{q}^2+f^2}=-\coth( f)\xrightarrow{f\rightarrow -\infty}1 . \\
\end{split}
\ee
Indeed we have recovered the known quench dynamics of the total negativity from the charge imbalance resolved ones, confirming our results.
\section{Numerical test}\label{section6}
\par In this section, we first review the numerical approach to the charged logarithmic negativity after a local joining quench (see \cite{Chen:2022gyy,Wen:2015qwa,Calabrese:2012nk}   for more details). Then we compare the CFT predictions with the numerical results.
\subsection{Numerical approach}
\begin{figure}
        \centering
        \subfloat
        {\includegraphics[width=8cm]{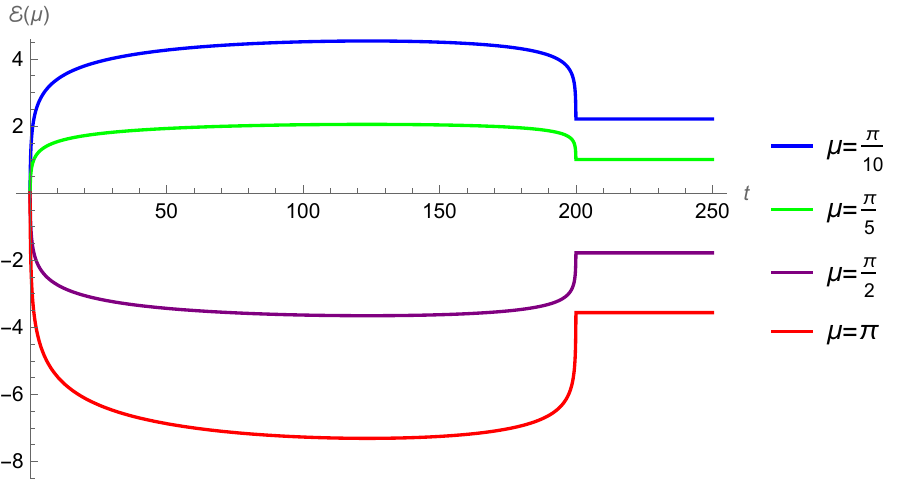}} \quad\quad
        {\includegraphics[width=8cm]{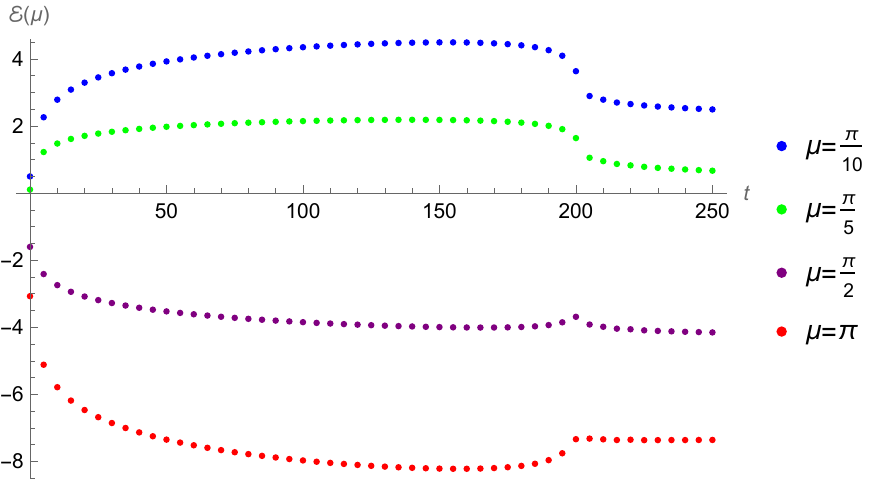}}
        \caption{The charged logarithmic negativity $\mE(\m)$ for two symmetric  adjacent finite intervals as a function of time $t$ for $\m=\frac{\pi}{10},\frac{\pi}{5},\frac{\pi}{2},\pi$, respectively. Here we choose $L=500, l=200$ and $\epsilon=0.04$. Left panel: CFT predictions. Right panel: numerical results in the complex harmonic chain. }
        \label{adjsym}
\end{figure}
\par To check our CFT predictions obtained in previous sections, we will consider the time evolution of the logarithmic negativity after a local joining quench in a complex harmonic chain which is the lattice version of our complex Klein-Gordon field theory. The Hamiltonian of the real harmonic chain is
\be
H_{RHC}=\sum_{j=1}^{L-1}\lt(\frac{1}{2M}p_j^2+\frac{M\o^2}{2}q_j^2+\frac{K}{2}(q_{j+1}-q_j)^2\rt),
\ee
where the Dirichlet boundary condition $q_0=q_L=p_0=p_L=0$ are imposed and variables $p_j$ and $q_j$ satisfy standard bosonic commutation relations $[q_i,q_j]=[p_i,p_j]=0$ and $[q_i,p_j]=\mathrm{i}\d_{ij}$. $L$ is the number of sites of the chain, $M$ is the mass scale, and $K$ is the nearest-neighbor coupling constant. The complex harmonic chain is equivalent to two decoupled real harmonic chains. In terms of the variables $q^{(1)},p^{(1)}$ and $q^{(2)},p^{(2)}$, the Hamiltonian of the complex harmonic chain can be written as
\be\label{HCHC}
H_{CHC}(p^{(1)}+\mathrm{i}p^{(2)},q^{(1)}+\mathrm{i}q^{(2)})=H_{RHC}(p^{(1)},q^{(1)})+H_{RHC}(p^{(2)},q^{(2)}).
\ee
For the Dirichlet boundary condition, the Fourier sine transform of the canonical variables are
\be
p_j=\sum_{k=1}^{L-1}  \widetilde{p}_k  \sqrt{\frac{2}{L}} \sin \left (   \frac{\pi k n}{L} \right ) , \quad
\widetilde{p}_k= \sum_{j=1}^{L-1} p_{j} \sqrt{\frac{2}{L}} \sin \left( \frac{\pi k n}{L} \right ),
\ee
where $j,k=1,\cdots, L-1$. For $q_j$, the Fourier sine transformation is defined similarly. In the momentum space, the Hamiltonian of the real harmonic chain is diagonal
\begin{align}
H_{RHC}=\sum_{k=1}^{L-1}\left ( \frac{1}{2M}  \widetilde{p}_k^2+\frac{M \omega_k^2}{2} \widetilde{q}_k^2 \right ),
\label{Eq: RHC}
\end{align}
where the dispersion relation reads \cite{Wen:2015qwa}
\be
\label{drr}
\omega_k=\sqrt{\omega_0^2 + \frac{4 K}{M} \sin (\frac{\pi k}{2L})^2}>\o_0.
\ee
Here $\o_0=0$ is well-defined while it is ill-defined in the periodic case. Now we introduce the creation and annihilation operators $a_k,a_k^{\dg}$ and $b_k,b_k^{\dg}$, satisfying $[a_k,a_{k'}^{\dg}]=\d_{kk'}$ and $[b_k,b_{k'}^{\dg}]=\d_{kk'}$. Then the Hamiltonian of the complex harmonic chain becomes
\be
H_{CHC}=\sum_{k=1}^{L-1}\o_k(a_k^{\dg}a_k+b_k^{\dg}b_k).
\ee
\begin{figure}
        \centering
        \subfloat
        {\includegraphics[width=8cm]{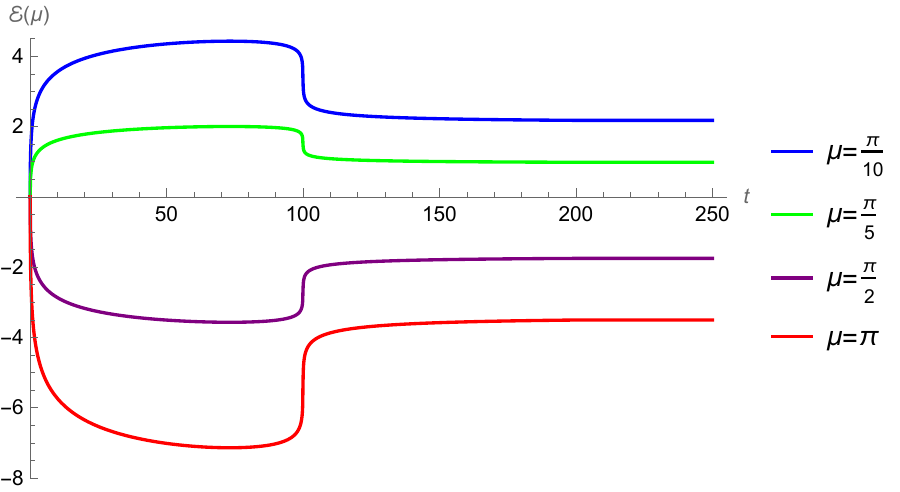}} \quad\quad
        {\includegraphics[width=8cm]{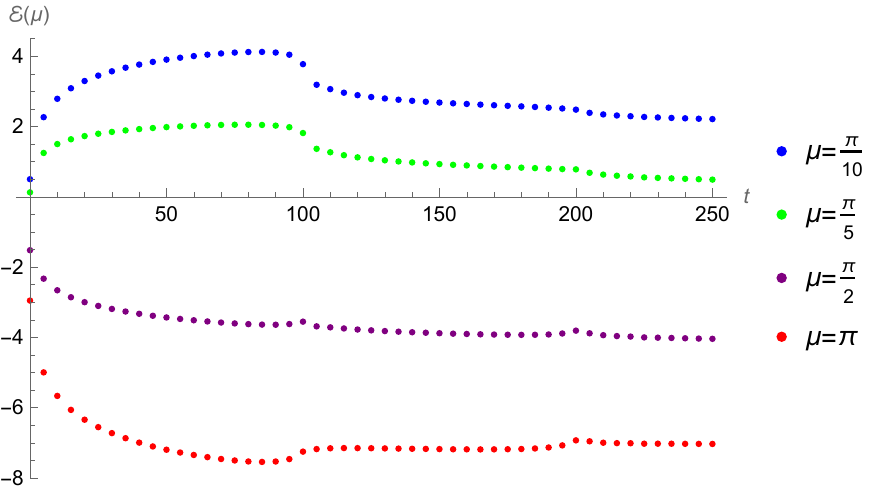}}
        \caption{The charged logarithmic negativity $\mE(\m)$ for two asymmetric  adjacent finite intervals as a function of time $t$ for $\m=\frac{\pi}{10},\frac{\pi}{5},\frac{\pi}{2},\pi$, respectively. Here we choose $L=500$, $l_1$=100, $l_2$=200 and $\epsilon=0.03$. Left panel: CFT predictions. Right panel: numerical results in the complex harmonic chain. }
        \label{adjasym}
\end{figure}
\par  The connection between the covariance matrix and the entanglement entropy was first derived  in Refs. \cite{Chung:2001zz,Peschel:2003,Peschel:2009}. Next, we use the correlation matrix method to obtain the charged logarithmic negativity. The covariance matrix of the real harmonic chain is constructed from two-point correlators of canonical variables
\be
\label{G(0)}
\G_{n,m}=\begin{pmatrix}
\mathbb{Q}_{n,m}&\mathbb{M}_{n,m}\\
\mathbb{M}_{n,m}^{\mathrm{T}}&\mathbb{P}_{n,m}
\end{pmatrix},
\ee
where
\be
\label{qpr}
\begin{split}
&\mathbb{Q}_{n,m}\equiv\bra{0}q_n q_m\ket{0}=\frac{1}{L}\sum_{k=1}^{L-1}\frac{1}{M\o_k}\sin(\frac{\pi kn}{L})\sin (\frac{\pi km}{L}),\\
&\mathbb{P}_{n,m}\equiv\bra{0}p_n p_m\ket{0}=\frac{1}{L}\sum_{k=1}^{L-1} M\o_k \sin(\frac{\pi k n}{L})\sin (\frac{\pi km}{L}),\\
&\mathbb{M}_{n,m}\equiv\bra{0}q_n p_m\ket{0}=\frac{\mathrm{i}}{2}\delta_{nm},
\end{split}
\ee
with $n,m=1,\cdots, L$. Using the Heisenberg equation of motion, $\dot{ \widetilde{q}}_k(t)=\frac{1}{M} \widetilde{p}_k(t)$ and $\dot{ \widetilde{p}}_k(t)=-M \omega_k^2 \widetilde{q}_k(t)$
, we have
\begin{equation}
\begin{split}
 \widetilde{q}_k(t)&=\frac{1}{\sqrt{M}} (\cos (\omega_k t)  \widetilde{q}_k(0) + \omega_k^{-1} \sin(\omega_k t) \widetilde{p}_k(0)), \\
 \widetilde{p}_k(t)&=\sqrt{M}(-\omega_k \sin(\omega_kt)  \widetilde{q}_k(0) + \cos(\omega_kt)  \widetilde{p}_k(0)).
\end{split}
\end{equation}
The time-dependent canonical variables in the real space are
\begin{equation}
\begin{split}
q_j(t)=&\frac{1}{\sqrt{M}}\sum_{k,m} \phi^*_k(n) \phi_k(m)\times\left (\cos(\omega_k t) q_m(0) + \omega_k^{-1} \sin(\omega_kt) p_m(0) \right ), \\
p_j(t)=&\sqrt{M}\sum_{k,m}\phi^*_k(n) \phi_k(m) \times\left (-\omega_k \sin(\omega_kt) q_m(0) + \cos(\omega_kt) p_m(0) \right ),
\end{split}
\end{equation}
\begin{figure}
        \centering
        \subfloat
        {\includegraphics[width=8cm]{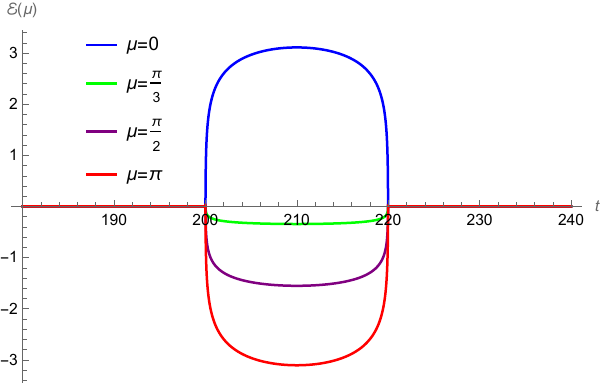}} \quad\quad
        {\includegraphics[width=8cm]{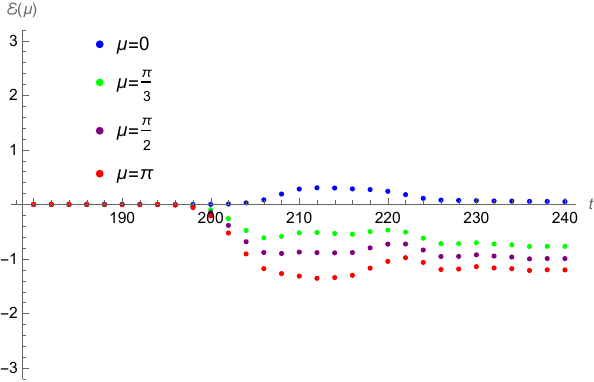}}
        \caption{The charged logarithmic negativity $\mE(\m)$ for two disjoint symmetric intervals as a function of time $t$ for $\m=0,\frac{\pi}{3},\frac{\pi}{2},\pi$, respectively. Here we choose $L=500, l=20, d=200$ and $\epsilon=0.02$. Left panel: CFT predictions. Right panel: numerical results in the complex harmonic chain. }
        \label{disjointsym}
\end{figure}\\
where
\begin{align}
 \phi_k(m)&= \sqrt{\frac{2}{L}}\sin  (\frac{\pi  k m}{L}).
\end{align}
The time evolution of the covariance matrix can be obtained from
\begin{equation}\label{Eq: CMt}
\Gamma(t) = S (t) \Gamma (0) S(t)^\mathrm{T},
\end{equation}
where the evolution matrix $S(t)$ is given by
\begin{equation}\label{Eq: SM}
\begin{split}
S_{n,m} (t)=&\sum_{k} \phi^*_k(n) \phi_k(m)\left(\begin{array}{cc}  \frac{1}{\sqrt{M}} \cos(\omega_k t)  & \frac{1}{\sqrt{M}} \omega_k^{-1} \sin(\omega_k t)  \\
 -\sqrt{M}\omega_k \sin(\omega_k t)  & \sqrt{M}\cos(\omega_k t) \end{array}\right)\\
=&\sum_{k} \frac{2}{L} \sin{\frac{\pi k n}{L}} \sin{\frac{\pi k m}{L}} \left(\begin{array}{cc}  \frac{1}{\sqrt{M}} \cos(\omega_k t)  & \frac{1}{\sqrt{M}} \omega_k^{-1} \sin(\omega_k t)  \\-\sqrt{M}\omega_k \sin(\omega_k t)  & \sqrt{M}\cos(\omega_k t) \end{array}\right).
\end{split}
\end{equation}
\par We first consider time evolution of covariance matrix $\Gamma_A(t)$ and symplectic matrix $J_A$ associated with the subsystem $A$
\be
\G_A(t)=\begin{pmatrix}
\mathbb{Q}_A(t)&\mathbb{M}_A(t)\\
\mathbb{M}_A(t)^{\mathrm{T}}&\mathbb{P}_A(t)
\end{pmatrix},\qquad
J_A=\begin{pmatrix}
\mathbb{O}&\mathbb{I}\\
-\mathbb{I}&\mathbb{O}
\end{pmatrix},\qquad
\ee
where $\mathbb{O}$ is zero matrix and $\mathbb{I}$ is identity matrix acting on the subsystem $A$. We denote the spectrum of $\mathrm{i}J_A\G_A(t)$ by $\{\pm\l_1(t),\cdots,\pm\l_{l_A}(t)\}$.
\par We introduce the Fock space basis
$\ket{\mathbf{n}}\equiv\otimes_{j=1}^l\ket{n_j}$, defined by products of eigenstates of the number operator in the subsystem $A$, the reduced density matrix of the subsystem $A$ can be written as
\be
\label{rhoAt}
\rho_A(t)=\sum_{\mathbf{n}}\prod_{j=1}^{l}\frac{1}{\l_j(t)+1/2}\lt(\frac{\l_j(t)-1/2}{\l_j(t)+1/2}\rt)^{n_j}\ket{\mathbf{n}}\bra{\mathbf{n}}.
\ee
which was first derived in Ref.\cite{Murciano:2020} for bosons at $t=0$.
\par In the Fock basis $\{\ket{\mathbf{n}}\}$, $Q_2^{T_2}=Q_2$ and the operator $\mQ_A=Q_1-Q^{T_2}_2=Q_1-Q_2$ becomes exactly the charge imbalance operator. We consider the partial transposition of $\G_A(t)$ with respect to $A_2$
\be
\G^{T_2}_A(t)=\begin{pmatrix}
\mathbb{I}_{A_1}&\mathbb{O}\\
\mathbb{O}&\mathbb{R}_{A_2}
\end{pmatrix}\G_A(t)\begin{pmatrix}
\mathbb{I}_{A_1}&\mathbb{O}\\
\mathbb{O}&\mathbb{R}_{A_2}
\end{pmatrix},
\ee
where $\mathbb{R}_{A_2}$ is the $l_2\times l_2$ diagonal matrix with elements $(\mathbb{R}_{A_2})_{nm}=\d_{nm}\begin{pmatrix}1&0\\0&-1\end{pmatrix}$.
\par For the complex harmonic chain, the charged R\'enyi negativity factorizes as
\be
N_{n}(\m)=\Tr[(\rho^{T_2}_A)^ne^{\mathrm{i}\m \mQ_A^a}]\times\Tr[(\rho^{T_2}_A)^ne^{-\mathrm{i}\m \mQ_A^b}].
\ee
\par If we denote the eigenvalues of $\mathrm{i}J_A\G_A(t)^{T_2}$ by $\{\pm\s_1(t),\pm\s_2(t),\cdots,\pm\s_{l_A}(t)\}$,
then the charged  R\'enyi logarithmic negativity is given by \cite{Chen:2022gyy}
\be
\mathcal{E}_n(\m)=-2\sum_{j=1}^{l_A}\log\Big|\lt(\s_j(t)+\frac12\rt)^n-e^{\mathrm{i}\m}\lt(\s_j(t)-\frac12\rt)^n\Big|.
\ee
For $n=1$, we have
\be
\mathcal{E}_1(\m)=-2\sum_{j=1}^{l_A}\log\Big|\lt(\s_j(t)+\frac12\rt)-e^{\mathrm{i}\m}\lt(\s_j(t)-\frac12\rt)\Big|.
\ee
The charged logarithmic negativity is given by
\be
\mathcal{E}(\m)=-2\sum_{j=1}^{l_A}\log\Big||\s_j(t)+\frac12|-e^{\mathrm{i}\m}|\s_j(t)-\frac12|\Big|.
\ee
\begin{figure}
        \centering
        \subfloat
        {\includegraphics[scale=.71]{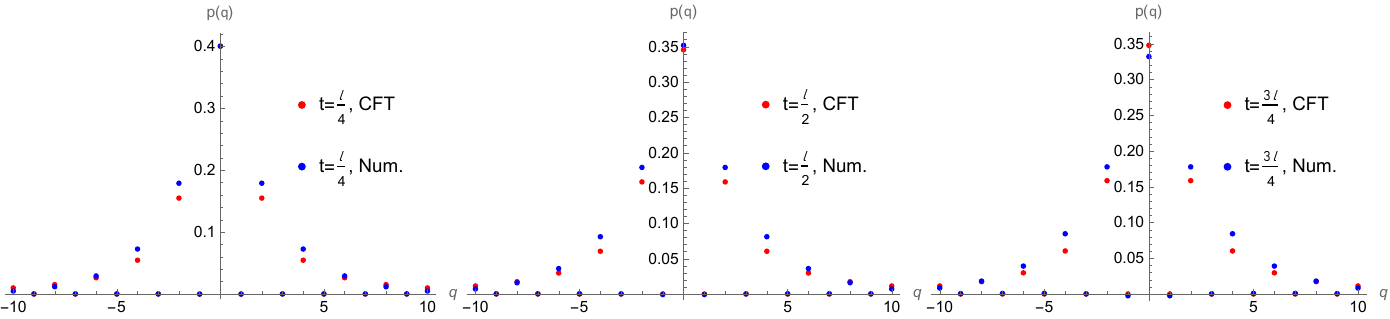}}
        \caption{The probability $p(\mathrm{q})$ for two adjacent symmetric finite intervals as a function of $\mathrm{q}$ for $t= \frac{l}{4},\frac{l}{2},\frac{3l}{4}$, respectively. Here we choose $ L=500, l=200$ and $\epsilon=3.2$. Red dots: CFT predictions. Blue dots: numerical results in the complex harmonic chain.}
        \label{pq}
\end{figure}
\par At time $t<0$, we consider two disconnected harmonic chains  with equal number of sites $L$ under the Dirichlet boundary condition with each harmonic chain initially prepared in its ground state. At time $t=0$, two disconnected chains are joined together as one harmonic chain with the number of sites $2L$ under the Dirichlet boundary condition. The evolution matrix in the situation is
\begin{equation}
\begin{split}
S_{n,m} (t)=&\frac{1}{L} \sum_{k=1}^{2L-1}   \sin{\frac{\pi k n}{2L}} \sin{\frac{\pi k m}{2L}} \left(\begin{array}{cc}  \frac{1}{\sqrt{M}} \cos(\O_k t)  & \frac{1}{\sqrt{M}} \O_k^{-1} \sin (\O_k t)  \\
 -\sqrt{M}\O_k \sin(\O_k t)  & \sqrt{M}\cos(\O_k t) \end{array}\right),
\end{split}
\end{equation}
where $\O_k=\sqrt{\o_0^2+\frac{4K}{M} \sin{(\frac{\pi k}{4L})^2}}$. We calculate the charged logarithmic negativity by setting $M=K=1$, $\o_0=0$. We choose the total length of harmonic chain $L=500$. The numerical data of quench dynamics of the charged logarithmic negativity are shown in Fig.~\ref{adjsym}, Fig.~\ref{adjasym} and  Fig.~\ref{disjointsym}. The numerical data of probability $p(\mathrm{q})$ is shown in Fig.~\ref{pq}. By comparing the CFT predictions with the numerical results obtained in the harmonic chain, we find that the main features agree well.
\subsection{Comparison between CFT predictions and numerical results}
\par
\begin{figure}
\centering
\includegraphics[scale=0.7]{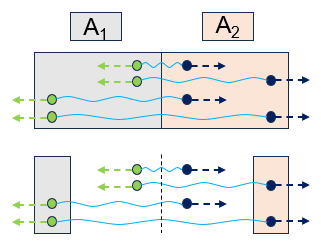}
\caption{The classification of entangled pairs on charged logarithmic negativity $\mE(\m)$ of two adjacent intervals and two disjoint intervals in the lattice model. For two adjacent intervals cases, all four kinds of entangled pairs contribute to the charged logarithmic negativity. For two disjoint intervals cases, however, only the fast-fast pairs can contribute to the charged logarithmic negativity near $t\approx d$.}
\label{EP}
\end{figure}

 The quasi-particle picture of entanglement negativity after a local joining quench has been proposed in Ref.\cite{Wen:2015qwa}. In CFT all the quasi-particles propagate at the speed of light $v_c$, while in the lattice model the situation is different. The dispersion relation in Eq.~(\ref{drr}) can be written as
\be
\omega(p)=\sqrt{\omega^2 + 4  \sin (\frac{p}{2})^2},\quad p\equiv \frac{\pi k}{L},
\ee
from which it is easy to calculate the group velocities
\be
v \equiv \frac{\partial \o(p) }{\partial p}=\frac{\sin (p)}{\sqrt{\o^2+4\sin^2 (p/2)}}.
\ee
It's easy to see that not all the quasi-particles have the same group velocities. In detail, the high-energy quasi-particles have the group velocities $v<v_c$. Following Ref. \cite{Wen:2015qwa}, as shown in Fig.~\ref{EP}, we can divide the entangled pairs into four classes according to their group velocities $(v_L,v_R)$.
\be
(v_L,v_R)\simeq
\left\{
\begin{aligned}
&(-v,v),&\text{slow-slow pair}\\
&(-v,v_c),&\text{slow-fast pair}\\
&(-v_c,v),&\text{fast-slow pair}\\
&(-v_c,v_c),&\text{fast-fast pair}\\
\end{aligned}
\right.
\ee
\par Although the main features of CFT predictions agree well with numerical results, we also observe that there are some disagreements between the two approaches. In the two adjacent intervals case, as shown in Fig.~\ref{adjsym} and Fig.~\ref{adjasym}, the charged logarithmic negativity obtained in the harmonic chain arrives at the ground state value gradually and the CFT results drop to the ground state value very suddenly.
The reason is that there are only the contributions of fast-fast pairs in CFT. However, in the lattice model the slow-slow pairs still contribute to $\mathcal{E}(\m)$ even for $t>l$ in the symmetric case and  $t>l_1$ in the asymmetric case. We also notice that the absolute value of $\mE(\m)$ of CFT results are smaller than the numerical results for $\m=\frac\pi 2, \pi$ during $t>l$ in the symmetric  case and $t>l_1$ in the asymmetric case. For two disjoint intervals case, as shown in Fig.~\ref{disjointsym}, the first disagreement is that the concrete values of the charged logarithmic negativity between CFT results and numerical results, namely, the numerical results are much smaller than the CFT results. By using the quasi-particle picture in Fig.~\ref{EP} again,  we can explain this phenomenon. For two adjacent intervals, all four kinds of entangled pairs contribute to the charged logarithmic negativity. In the two disjoint intervals case, however, only the fast-fast pairs can contribute to the charged logarithmic negativity during $d<t<d+l$ in the symmetric case, and the other three kinds of entangled pairs do not make any contribution at all. Thus the CFT results are much larger than the numerical results for disjoint case. However, there is disagreement during  $t>d+l$: The numerical results of $\mE(\m)$ are non-zero while in CFT they vanish identically, which is due to the contribution of slow-slow modes in the lattice model. 
\section{Conclusion}\label{section7}
\par In this paper, we studied the dynamics of the charge imbalance resolved entanglement negativity after a local joining quench in $1+1$ dimensional free complex boson CFT. Firstly, we use the conformal mapping approach of local joining quenches to calculate the time evolution the charge-resolved negativity. In this quench protocol, the entangled pairs are generated at the joining point and then propagate freely through the system. Secondly, we compare our CFT predictions with the numerical results for quench dynamics of charged logarithmic negativity and probability. Finally, we explained the phenomenon based on the quasi-particle picture.
\par It will be an interesting problem to see how our results are changed in finite temperatures in the same quench protocol discussed in this paper. More interestingly, one could consider the non-equilibrium evolution of charge resolved entanglement where two CFTs are initially prepared at different temperatures suddenly joined together. In this setting, a non-equilibrium steady state should exist at late times \cite{Hoogeveen:2014bqa}.
\section*{Acknowledgments}
This work was supported  by the National Natural Science Foundation of China, Grant No.\ 12005081.

\end{document}